# Energy Efficient Location Aided Routing Protocol for Wireless MANETs

Mohammad A. Mikki
*Computer Engineering Department*
*IUG*
Gaza, Palestine
mmikki@iugaza.edu.ps

*Abstract*— A Mobile Ad-Hoc Network (MANET) is a collection of wireless mobile nodes forming a temporary network without using any centralized access point, infrastructure, or centralized administration.

In this paper we introduce an Energy Efficient Location Aided Routing (EELAR) Protocol for MANETs that is based on the Location Aided Routing (LAR). EELAR makes significant reduction in the energy consumption of the mobile nodes batteries by limiting the area of discovering a new route to a smaller zone. Thus, control packets overhead is significantly reduced. In EELAR a reference wireless base station is used and the network's circular area centered at the base station is divided into six equal sub-areas. At route discovery instead of flooding control packets to the whole network area, they are flooded to only the sub-area of the destination mobile node. The base station stores locations of the mobile nodes in a position table. To show the efficiency of the proposed protocol we present simulations using NS-2. Simulation results show that EELAR protocol makes an improvement in control packet overhead and delivery ratio compared to AODV, LAR, and DSR protocols.

I. INTRODUCTION

A mobile ad hoc network (MANET) consists of a group of mobile nodes (MNs) that communicate with each other without the presence of infrastructure. MANETs are used in disaster recovery, rescue operations, military communication and many other applications. In order to provide communication throughout the network, the mobile nodes must cooperate to handle network functions, such as packet routing. The wireless mobile hosts communicate in a multi-hop fashion. In multi-hop wireless ad-hoc networks, designing energy-efficient routing protocols is critical since nodes have very limited energy, computing power and communication capabilities. For such protocols to scale to larger ad-hoc networks, localized algorithms need to be proposed that completely depend on local information. The key design challenge is to derive the required global properties based on these localized algorithms.

In ad hoc networks, the routing protocols are divided into three categories: Proactive, Reactive and Hybrid. In Proactive routing protocols, each MN maintains a routing table where control packets are broadcasted periodically within the whole network. This means that the routes to destination MNs are computed at a regular time before establishing the connection from source to destination. When a source MN wants to send data to a destination MN, it searches the routing table to find a destination MN match. The advantage of such a method is that the route is already known. But the disadvantage is that the control packets overhead is large since they are sent periodically to maintain all routes although not all routes will be necessarily used. Thus, the limited network bandwidth is consumed by control overhead. An example of proactive routing protocol is DSDV [9].

In Reactive routing protocols, the routes are discovered only when the source MN needs to transmit data packets. Thus, the control packets are broadcasted just when there are data to be transmitted. So, the broadcast overhead is reduced. In these protocols, there are two phases to establish routes to destination. These two phases are route discovery and route maintenance. Since the nature of the ad hoc network is highly mobile, the topology of the network is changed often. When the route to destination is broken, the route maintenance phase is started to keep route available. This method suffers from large end to end delay to have route available before sending data packets in large networks. An example of reactive routing protocol is DSR [5].

Hybrid routing protocols include the advantages of both proactive and reactive protocols. Each MN defines two zones: the inside zone and the outside zone. Each node maintains a neighbor table with n MN hops. These MNs are considered to be in the inside zone of the node. Thus, the hybrid protocols act as proactive protocols in the inside zone and reactive protocols in the outside zone. Each node periodically broadcasts control packets in the inside zone to build a routing table for all MNs in the inside zone. When a node wishes to send data to a destination node that resides in the outside zone, it uses a reactive protocol. Thus, a route discovery phase is invoked to establish the route to the destination MN. An example of Hybrid routing protocols is ZRP [14].

When the routing protocol does not use the location information of the mobile node, then the routing is topology-based routing protocol. If the position information is used in the routing protocol, then the routing is position-based routing protocol [15], [16]. There are two methods of forwarding data packets in position-based routing: greedy forwarding and directional flooding [23]. In greedy forwarding, the next hop node is the closest in distance to destination. Greedy Perimeter Stateless Routing Protocol (GPSR) uses the greedy forwarding [6]. In the directional flooding [19], the source node floods data packets in a geographical area towards the direction of the destination node. Location Aided Routing (LAR) uses directional forwarding flooding [1], [19].

In the position-based routing protocols, an MN uses a directional antenna or GPS system to estimate its (x, y) position. If GPS is used, every node knows it's (x, y) position assuming z = 0. Fig. 1 shows two mobile nodes with their positions determined using GPS. The positions of the two mobile nodes in Fig. 1 are (x1, y1) and (x2, y2) respectively. Using Fig. 1, the distance d between the two MNs is calculated using (1). The angle θ is defined as shown in Fig. 1 and is calculated using (2).

$$d = \sqrt{(x2 - x1)^2 + (y2 - y1)^2} \qquad (1)$$

$$\theta = \tan^{-1}\frac{(y2-y1)}{(x2-x1)} \qquad (2)$$

When directional antennas are used, the distance between two MNs and Angle of Arrival (AoA) are estimated according to the directional arrival. The strength of the signal is used to estimate the distance between two nodes and the estimate of θ is obtained from the Angle of Arrival (AoA) [12], [13].

The rest of the paper is organized as follows: Section II presents related work. Section III presents EELAR approach. Section IV validates the proposed approach. Finally, section V concludes the paper.

## II. RELATED WORK

In this section we present some of the most important routing protocols used in wireless mobile ad hoc networks.

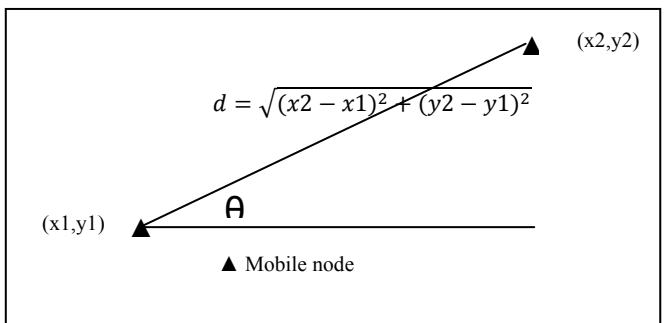

Figure 1. Position-based routing protocol that uses GPS to determine mobile nodes (x,y) positions

The Dynamic Source Routing (DSR) protocol is a simple and efficient routing protocol designed specifically for use in multi-hop wireless ad hoc networks of mobile nodes. DSR allows the network to be completely self-organizing and self-configuring, without the need for any existing network infrastructure or administration. The protocol is composed of the two mechanisms: route discovery and route maintenance, which work together to allow nodes to discover and maintain source routes to arbitrary destinations in the ad hoc network [5]. The DSR protocol is triggered by a packet generated at the sending node for a destination node whose IP address is (or can be) known to the sending node. When a node has a packet to send to a destination it first checks its cache if a path to the destination is already known. If the path is not available then the route discovery mechanism is initiated. Route Discovery allows any host in the ad hoc network to dynamically discover a route to any other host in the ad hoc network. The Route Maintenance procedure monitors the operation of the routes and informs the sender of any routing errors. Route maintenance is required by all routing protocols, especially the ones for MANETs due to very high probability of routes being lost [11]. The use of source routing allows packet routing to be trivially loop-free, avoids the need for up-to-date routing information in the intermediate nodes through which packets are forwarded, and allows nodes forwarding or overhearing packets to cache the routing information in them for their own future use. All aspects of the protocol operate entirely on-demand, allowing the routing packet overhead of DSR to scale automatically to only that needed to react to changes in the routes currently in use [17].

The Multipoint Relays (MPR) technique efficiently fulfills the flooding function in wireless networks. It is a technique to reduce the number of redundant re-transmission while diffusing a flooding packet throughout the entire network. Each node N in the network selects some neighbors as its Multipoint Relays (MPR). Only these neighbors will retransmit the flooding packets broadcasted by node N. These nodes called 2-hop neighbors whose distance to N is 2 hops. The MPR selection algorithm should guarantee that the flooding packets from N will be received by all its 2-hop neighbors after re-broadcast of N's MPRs.

Location-Aided Routing (LAR) protocol is an approach that decreases overhead of route discovery by utilizing location information of mobile hosts. Such location information may be obtained using the global positioning system (GPS) [1], [6], [7], [8], [19]. LAR uses two flooding regions, the forwarded region and the expected region. LAR protocol uses location information to reduce the search space for a desired route. Limiting the search space results in fewer route discovery messages [1], [19]. When a source node wants to send data packets to a destination, the source node first should get the position of the destination mobile node by contacting a location service which is responsible of mobile nodes positions. This causes a connection and tracking problems [8], [10]. Two different LAR algorithms have been presented in [19]: LAR scheme 1 and LAR scheme 2. LAR scheme 1 uses expected location of the destination (so-called expected zone) at the time of route discovery in order to determine the request zone. The request zone used in LAR scheme 1 is the smallest rectangle including current location of the source and the expected zone for the destination. The sides of the rectangular request zone are parallel to the X and Y axes. When a source needs a route discovery phase for a destination, it includes the four corners of the request zone with the route request message transmitted. Any intermediate nodes receiving the route request then make a decision whether to forward it or not, by using this explicitly specified request zone. Note that the request zone in the basic LAR scheme 1 is not modified by any intermediate nodes. On the other hand, LAR scheme 2 uses distance from the previous location of the destination as a parameter for defining the request zone. Thus, any intermediate node J receiving the route request forwards it if J is closer to or not much farther from the destination's previous location than node I transmitting the request packet to J. Therefore, the implicit request zone of LAR scheme 2 becomes adapted as the route request packet is propagated to various nodes.

AODV [22] protocol is a distance vector routing protocol that operates on-demand. There are no periodic routing table exchanges. Routes are only set up when a node wants to communicate with some other node. Only nodes that lie on the path between the two end nodes keep information about the route. When a node wishes to communicate with a destination node for which it has no routing information, it initiates route discovery. The aim of route discovery is to set up a bidirectional route from the source to the destination. Route discovery works by flooding the network with route request (RREQ) packets. Each node that receives the RREQ looks in its routing table to see if it is the destination or if it has a fresh enough route to the destination. If it does, it sends a unicast route reply (RREP) message back to the source, otherwise it rebroadcasts the RREQ. The RREP is routed back on a temporary reverse route that was created by the RREQ. Each node keeps track of its local connectivity, i.e., its neighbors. This is performed either by using periodic exchange of HELLO messages, or by using feedback from the link layer upon unsuccessful transmission. If a route in the ad hoc network is broken then some node along this route will detect that the next hop router is unreachable based on its local connectivity management. If this node has any active neighbors that depend on the broken link, it will propagate route error (RERR) messages to all of them. A node that receives a RERR will do the same check and if necessary propagate the RERR further in order to inform all nodes concerned.

### III. ENERGY EFFICIENT LOCATION AIDED ROUTING PROTOCOL APPROACH

This section presents our proposed Energy Efficient Location Aided Routing (EELAR) protocol approach. The proposed protocol is a modification to the ad hoc routing protocol LAR [1], [19]. EELAR utilizes location information of mobile nodes with the goal of decreasing routing-related overhead in mobile and ad hoc networks. It uses location information of the mobile nodes to limit the search for a new route to a smaller area of the ad hoc network which results in a significant reduction in the number of routing messages and therefore the energy consumption of the mobile nodes batteries is decreased significantly. In order to reduce the control overhead due to broadcast storm in the network when control packets are flooded into whole network (as in DSR protocol for example) EELAR uses a wireless base station (BS) that covers all MNs in the network. BS divides the network into six areas as shown in Fig. 2.

In order for BS to efficiently route packets among MNs, it keeps a Position Table (PT) that stores locations of all MNs. PT is built by BS through broadcasting small BEACON packets to all MNs in the network. MNs local positions are estimated from directional antennas, the distance between the MN and BS is estimated using the strength of the signal from MN to BS, and the angle of arrival (AoA); θ (which is the angle of the mobile node from which the packet arrives to BS) is estimated using directional antenna of the MN. Based on the AoA, BS can determine the network area in which each MN is located.

Table I shows how θ decides the area ID of each MN. When a source MN needs to transmit data, it first queries BS about the area id of the destination MN, then data packets are flooded into that area only. The use of location information of the destination mobile node limits the search for a new route to one of the six areas of the ad hoc network.

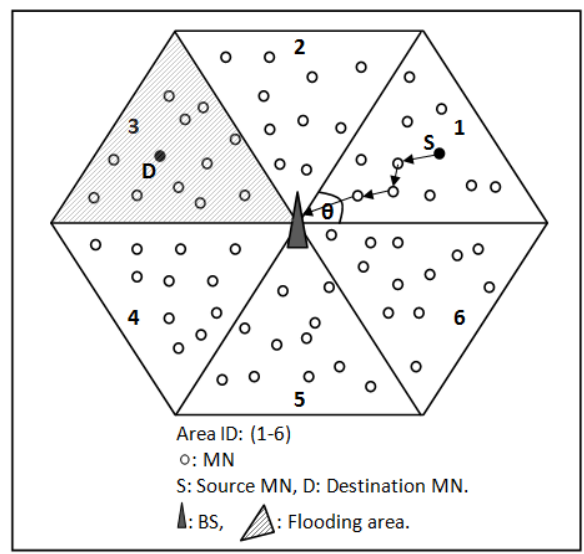

Figure 2. The definition of the six areas in EELAR

TABLE I. THE DEFINITION OF THE SIX NETWORK AREAS IN EELAR BASED ON θ

| Area ID | Range of angle θ |
|---|---|
| 1 | $0 \leq \theta < \pi/3$ |
| 2 | $\pi/3 \leq \theta < 2\pi/3$ |
| 3 | $2\pi/3 \leq \theta < \pi$ |
| 4 | $\pi \leq \theta < 4\pi/3$ |
| 5 | $4\pi/3 \leq \theta < 5\pi/3$ |
| 6 | $5\pi/3 \leq \theta < 2\pi$ |

Fig. 3 shows the pseudo code of EELAR. As Fig. 3 shows, the algorithm is multithreaded. First, it creates a thread that executes BuildUpdatePositionTable which builds and updates the PT in BS. Then, EELAR executes an infinite loop. In this loop, whenever a new mobile node enters the network area of BS then BuildUpdatePositionTable procedure is called so that the new mobile node will report its position to BS and hence, its position is included in the PT in BS. When a source mobile node S wants to send data packets to a destination mobile node D, EELAR creates a new thread that executes DataTransmission procedure. Multiple pairs of mobile nodes could communicate in parallel using parallel threads.

Fig. 4 shows the pseudo code of BuildUpdatePositionTable procedure. As Fig. 4 shows, BuildUpdatePositionTable procedure starts by handling the case when a mobile node A enters the network range of BS. A uses its location estimation method to determine its (x, y) position. A sends a broadcast message (PosReq message) that contains its position. PosReq message is a request to join the network of BS. PosReq contains the location of A. When BS receives this message it updates its PT. BS determines A's angle θ; distance d between A and BS; and classifies A as belonging to one of the six network areas. Then, BS replies with ID Reply message (IDRp message) to A that contains the area ID of A, hence A knows its area ID. Then, BuildUpdatePositionTable continues where BS periodically broadcasts BEACON packets to all MNs in the network in order to build PT that contains the network area ID of each MN that resides within the transmission range of BS. This scenario is repeated between BS and all MNs periodically as long as the mobile nodes are still in this network. When a mobile node stops sending the broadcast packet (PosReq) then it is marked unreachable by BS after a timer T expires.

Fig. 5 shows the pseudo code of DataTransmission procedure. DataTransmission procedure is called by EELAR when a source mobile node S sends data packets to a destination mobile node D. As Fig. 5 shows, first, S requests from BS to initiate a route discovery to node D by sending a DstPosReq (destination position request) packet to BS that requests the position information of D. BS checks if the position of D in PT is out of date, if so BS sends a small BEACON message to node D requesting its new location information to avoid out of date location information and updates its PT. Then, BS searches its position table for the area ID of D. When BS determines the area ID of D, it sends back DstIDRp (Destination ID Reply) packet to S containing the network area ID of D. If the BS determines that S and D are not in the same area then BS sends a control packet to S indicating that the data flow will be through BS, so each data packet from S to D will contain a "toBS" flag in the header forcing all nodes in S's area to drop these packets and not to handle them. Then, BS forwards data packets from node S to the area where D belongs only. When the source node S wants to transmit data to node D and BS determined that S and D are in the same network area, then BS will reply with a packet which indicates that the data flow will be done within the network area of node S and not through BS. This frees BS from being involved in the communication between S and D and BS will not be a performance bottleneck. Then node S floods its own area with data packets that are directed to D. If node B (which is in the same area as node S) receives a data packet directed to D and originating from S (B may receive this packet from any node in same area of S) then it measures the distance between itself and D and compares it with the distance between S and D. If B's distance is less than the S's distance then B will forward the packet. Otherwise, it will drop it.

```
algorithm EELAR ( ) {
  Thread (BuildUpdatePositionTable); // create a thread
  // that executes BuildUpdatePositionTable procedure
  while (1) {
    if ( a mobile node enters network area  of the
        base station)
        Thread (BuildUpdatePositionTable)
    if (source mobile node wants to send data
        to a destination mobile node)
        Thread (DataTransmission); // create a thread that
        // executes DataTransmission procedure
  } // end while
} // end EELAR
```
Figure 3. EELAR pseudo code

```
procedure BuildUpdatePositionTable ( ) // build and
    // update position table in BS
Input: mobile node A; base station X; {
    Control packet PosReq; // position request
    // message containing x, y coordinates
    if (node A enters network area controlled by X){
        A sends PosReq  to X;
        X: addPositionTable ( A, x,y);
        X: sends IDRp to A containing area ID of A;
    }//end if
    Repeat every time T
        X sends BEACON message to A;
        A sends PosReq to X;
        X: UpdatePositionTable (A, x, y );
    until valid timer expires
    X marks node A unreachable
} // end BuildUpdatePositionTable
```
Figure 4. BuildUpdatePositionTable pseudo code

```
procedure DataTransmission ( ) //  a source mobile node
// S sends data to destination mobile node D
Input: Source node S, destination node D,
       base station X;
{
   // S initiates data transmission to D
   // S requests X to initiate routing discovery
   S sends DstPosReq to BS;
   X checks PT for position of D;
   if (position of D in PT is out of date){
      X sends BEACON message to D;
      D sends PosReq to X;
      X: UpdatePositionTable (A, x, y );
   } // end if
   X searches PT for position of D;
   X sends DstIDRp to S; // message contains
   // area ID of D
   if (isNotIntheSameArea (S, D) ) {
      S sets toBS flag in header of all packets to D;
      //  nodes in same area as S will drop the packet
      S sends data to X;
      X routes data to D; // BS floods message to
      // area of D
   } // end if
   else {
      // S floods message to its own area
      S sends data to same area nodes
      for each node B in S's area network  {
         if (distance (B,D) < distance (S, D) ) {
            B forwards this packet;
         else
            B drops this packet;
      } // end for
   } //end else
} // end procedure
```

Figure 5.  DataTransmission pseudo code

The benefit DataTransmission procedure is to make the amount of data that can be transmitted and received at time t more than the available bandwidth of BS through not involving BS with data transmission when this data transmission is between nodes that are in the same area.

## IV.  EXPERIMENTAL RESULTS

In order to validate the proposed protocol and show its efficiency we present simulations using network simulator version 2 (NS-2). NS-2 is a very popular network simulation tool. It uses C language for protocol definition and TCL scripting for building the simulation scenarios [21]. The simulation environment settings used in the experiments are shown in Table II. The simulation duration is 500 seconds and the  network area is 1500 meter x 1500 meter that includes variable number of mobile nodes ranging from 50 to 250. A Constant Bit Rate (CBR) is generated as a data traffic pattern at a rate of 2 packets per second, and 20% of the mobile nodes are selected randomly as CBR sources. The scenario of nodes mobility is generated randomly based on random way point model [20] where a mobile node moves to a new position and pauses there for time period between 0 to 3 seconds, then it move to another position.

TABLE II.    NS2 simulation environment settings

| Parameter | Setting Value |
| --- | --- |
| Simulation duration | 500 sec |
| Network area | 1500 m x 1500  m |
| Number of  mobile nodes | 50,100,150,200,250 |
| Mobility model | Random way point model |
| Pause time | 0 to 3 sec |
| Node transmission range | 250 m |
| Data packet size | 512 bytes |
| Number of CBR sources | 20% of MNs |
| CBR rate | 2 packets per second |
| Mobile node speed | 5 to 30 m/s |

We compare performance of EELAR with AODV, LAR, and DSR which are well known routing protocols in MANETs.  The measured performance metrics are control overhead and the data packets delivery ratio. The control overhead is the number of control packets divided by the number of delivered data packets in the network, and the data packets delivery ratio is the number of received data packets divided by the total number of sent data packets.

In the first experiment we measure the control overhead in the network of the four protocols as a function of the average speed of mobile nodes. The number of MNs in the network was set to 100 and the average speed of MNs was varied from 5 to 30 m/s. The result is shown in Fig. 6. As the figure shows, for all compared protocols the overhead increases slightly as the average speed of MNs increases. In addition, EELAR protocol has the smallest control overhead among the four compared protocols. LAR has the second smallest control overhead, AODV has the third smallest control overhead, and DSR has the worst control overhead. The justification for the small control overhead in EELAR compared to the rest of protocols is that control packets used in discovering a new route are limited to a smaller zone.

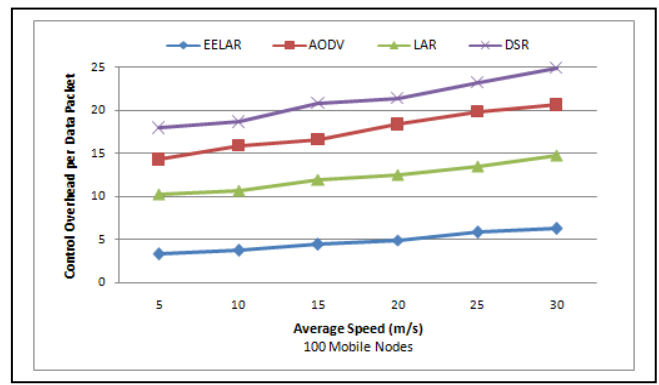

Figure 6.  Control overhead versus average speed

In the second experiment we measure the delivery ratio of data packets for the four compared protocols as a function of the average speed of mobile nodes. The number of MNs in the network was set to 100 and the average speed of MNs was varied from 5 to 30 m/s. The result is shown in Fig. 7. As the figure shows, for all compared protocols the data delivery ratio decreases slightly as the average speed of MNs increases. In addition, EELAR protocol has the highest delivery ratio of data packets among the four compared protocols. LAR has the second highest delivery ratio, AODV has the third highest delivery ratio, and DSR has the worst delivery ratio. As an explanation to the good delivery ratio in EELAR is that since control overhead is smaller (as shown in first experiment), the battery life of mobile nodes is longer, and hence routes are maintained for longer time. One reason for loss of data packets is the loss of the routes due to power shortage.

In the third experiment we measure the control overhead in the network of the four protocols as a function of the number of mobile nodes. The average speed of MNs was set to 15 m/s and the number of mobile nodes in the network was varied from 50 to 250 MNs. The result is shown in Fig. 8. The simulation results show that for all compared protocols the control overhead in the network is increased slightly as the node density of the network is increased. In addition, EELAR protocol has the smallest control overhead among the four compared protocols. LAR has the second smallest control overhead, AODV has the third smallest control overhead, and DSR has the worst control overhead. The justification of the improvement in control overhead in EELAR compared to the other three protocols is same as the justification presented in the case of the first experiment.

In the fourth experiment we measure the delivery ratio of data packets in the network of the four protocols as a function of the number of mobile nodes. . The average speed of MNs was set to 15 m/s and the number of mobile nodes in the network was varied from 50 to 250 MNs. The result is shown in Fig. 9. As the figure shows, for LAR, AODV and DSR the data delivery ratio increases very slightly and for EELAR the data delivery ratio remains the same as the number of MNs increases. In addition, EELAR protocol has the highest delivery ratio of data packets among the four compared protocols. Delivery ratio in EELAR never goes below 95%. LAR has the second highest delivery ratio, AODV has the third highest delivery ratio, and DSR has the worst delivery ratio. The justification of the improvement in delivery ratio in EELAR compared to the other three protocols is same as the justification presented in the case of the third experiment.

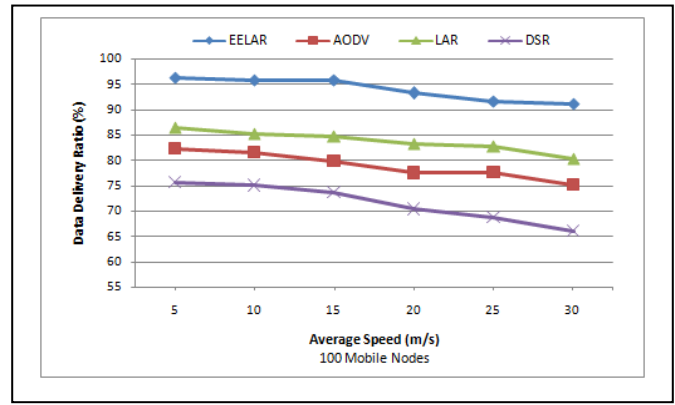

Figure 7. Data packets delivery ratio versus average speed

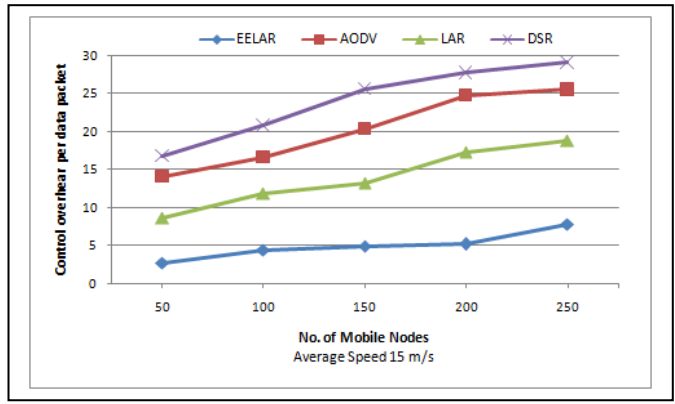

Figure 8. Control overhead versus number of MNs in the network

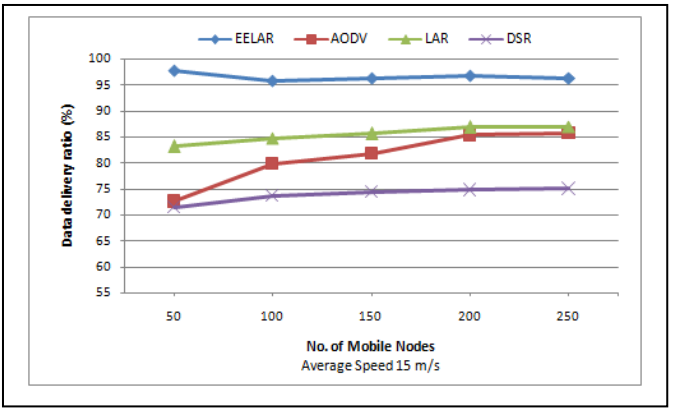

Figure 9. Data packets delivery ratio versus number of MNs in the network

In the last experiment we determine the optimal number of network areas that the network should be divided into, which produces the smallest control overhead. So we study the effect of varying number of network areas on control overhead in EELAR. Fig. 10 shows the result. In the experiment number of network areas was varied from 1 to 20, number of mobile nodes was set to 250 and the average speed was set to 15 m/s. As the figure shows the control overhead keeps decreasing as the number of network areas increases until this number reaches 6, then the control overhead starts increasing as we keep increasing number of network areas. This is explained as follows. For the control overhead decrease part: The idea of EELAR is to make significant reduction in control overhead by limiting the area of discovering a new route to a smaller zone. Thus, control overhead is reduced as number of areas increases. For the control overhead increase part: Increasing number of areas increases routes loss. When there is a very large number of areas and due to mobility of nodes, there is a higher probability that a node leaves its original area and enters a new area very quickly during a short period of time. Hence, in the case of larger number of areas when a source node initiates a transmission to a destination node, the possibility of lost routes during transmission period is higher than that in the case of smaller number of area. This leads to increased control overhead. This increased control overhead becomes worse as the number of areas keeps increasing.

Thus, our approach of dividing the network area into six sub-areas is not the optimal solution in all cases. There is a tradeoff between decreasing control overhead by increasing number of areas and route loss by increasing the number of network areas due to node mobility. This suggests that optimal number of network area is dependent on the nodes mobility.

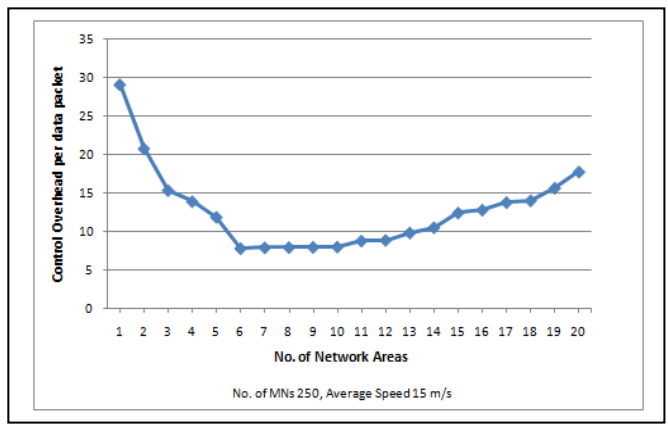

Figure 10. Control overhead in EELAR versus number of network areas

## V. CONCLUSION

This paper proposed an Energy Efficient Location Aided Routing Protocol (EELAR) that is an optimization to the Location Aided Routing (LAR). EELAR makes significant reduction in the energy consumption of the mobile nodes batteries through limiting the area of discovering a new route to a smaller zone. Thus, control packets overhead is significantly reduced and the mobile nodes life time is increased. To show the efficiency of the proposed protocol we presented simulations using NS-2. Simulation results show that our proposed EELAR protocol leads to an improvement in control overhead and delivery ratio compared to AODV, LAR, and DSR protocols.

In addition, simulation results show that there is a tradeoff between decreasing control overhead by increasing number of areas and increasing route loss by increasing the number of network areas due to node mobility. This suggests that optimal number of network area is dependent on the nodes mobility.

Suggestions for future work include developing a method to adaptively use one of the forwarding methods of the position-based routing protocol based on the surrounding environments, and dividing the network into a number of areas that varies dynamically based on the node mobility pattern.

ACKNOWLEDGMENT

The author wishes to acknowledge Mohamed B. AbuBaker, Shaaban A. Sahmoud and Mahmoud Alhabbash from the computer engineering department at IUG for their work, useful feedback, and comments during the preparation of this paper.

AUTHORS PROFILE

Mohammad A. Mikki is an Associate Professor of Parallel and Distributed Computing in the Electrical and Computer Engineering Department at IUG with about fifteen years of research, teaching, and consulting experience in various computer engineering disciplines. Dr. Mikki was the first chairman of the ECE department at IUG in the academic year of 1995-1996. He taught both graduate and undergraduate courses at the ECE department at IUG. In addition he taught several undergraduate courses at the College of Science and Technology, College of Education (currently Al-Aqsa University) and Al-Quds Open University. He was a visiting Professor at the Department of Electrical and Computer Engineering at University of Arizona in Tucson, Arizona **(USA)** during the academic year of 1999-2000. He was granted DAAD Study Visit Scholarship to Paderborn University in Paderborn in Germany from July 2002 to August 2002 from DAAD (German Academic Exchange Service). Dr. Mikki published about twenty publications in both journals and international conferences.

Dr. Mikki got both his Ph.D. and Master of Science in Computer Engineering from Department of Electrical and Computer Engineering in Syracuse University in Syracuse, New York, USA in December 1994 and May 1989 respectively. He also got his Bachelor of Science in Electrical Engineering from the Department of Electrical Engineering at BirZeit University in BirZeit in West Bank in August 1984.

Dr. Mikki got a graduate research assistantship from NPAC (North East Parallel Architecture Center) at Syracuse University in Syracuse in New York (USA) during the year of 1989-1990. He also got a Research Assistantship from the Department of Electrical and Computer Engineering at Syracuse University in Syracuse, in New York (USA) during the period of 1990-1994. He also received a Deanery of Scientific Research grants from IUG during the academic years of 01/02, 03/04, and 07-08. Dr. Mikki was a software consultant and programmer at Vertechs Software Solutions Inc in Syracuse in New York (USA) during the period from 1991 to 1994. He was also a software consultant at Computer Software Modeling and Analysis in Fayetteville in New York (USA) from January 1993 to March 1993.

Dr. Mikki got two funded projects from the European Union (EU): Mediterranean Virtual University (MVU) project from 2004 to 2006 and Open Distance Inter-university Synergies between Europe, Africa and Middle East (ODISEAME) project from 2002 to 2005.

Research Interests of Dr. Mikki include High Performance Parallel and Distributed Computing, Grid and Cluster Computing, Wireless and Mobile Networks, Modeling and Design of Digital Computer Systems, Internet Technology and Programming, Internet Performance Measurement Tools and Web-Based Learning